\documentclass[preprint,review,12pt]{elsarticle}

\usepackage{graphicx}

\usepackage{amssymb}
\usepackage{color}

\biboptions{sort&compress}

\journal{Computer Physics Communications}
\begin{document}

\begin{frontmatter}



\title{Hierarchies in Nucleation Transitions}


\author[mz]{Christoph Junghans}
\address[mz]{Max-Planck-Institut f\"ur Polymerforschung Mainz, Ackermannweg 10,\\ D-55128 Mainz, Germany}
\ead{junghans@mpip-mainz.mpg.de}
\author[lei]{Wolfhard Janke}
\address[lei]{Institut f\"ur Theoretische Physik and Centre for Theoretical
  Sciences (NTZ),\\ Universit\"at Leipzig, Postfach 100\,920,\\ D-04009 Leipzig, Germany}
\ead{Wolfhard.Janke@itp.uni-leipzig.de}
\author[ju]{Michael Bachmann}
\address[ju]{Soft Matter Systems Research Group, Institut f\"ur Festk\"orperforschung (IFF-2) and Institute for Advanced Simulation (IAS-2), Forschungszentrum J\"ulich,\\ D-52425 J\"ulich, Germany}
\ead{m.bachmann@fz-juelich.de}
\ead[url]{www.smsyslab.org}

\begin{abstract}
We discuss the hierarchy of subphase transitions in first-order-like nucleation
processes for an examplified aggregation transition of heteropolymers. We perform an
analysis of the microcanonical entropy, i.e., the density of states is considered 
as the central statistical system quantity since it connects system-specific 
entropic and energetic information in a natural and unique way.
\end{abstract}


\end{frontmatter}


\section{Introduction}
\label{intro}

The noncovalent cooperative effects in structure formation processes 
on mesoscopic scales let linear polymers be very 
interesting objects for studies of the statistical mechanics and thermodynamics
of nucleation processes, even on a fundamental level. Structural properties of
polymers can typically be 
well described by means of simple, coarse-grained models of beads and sticks (or springs) representing
the monomers and the covalent bonds between adjacent monomers in the chain, 
respectively. Contemporary, sophisticated
generalized-ensemble Monte Carlo simulation techniques 
as well as large-scale computational resources enable the 
precise and systematic analysis of thermodynamic properties of all structural phases of 
coarse-grained polymer models by means of computer simulations. Among the most efficient simulations methods are
multicanonical sampling~\cite{muca1,muca2}, replica-exchange techniques~\cite{huku1,geyer1}, 
and the Wang-Landau method~\cite{wl1}.

The high precision of the numerical data for quantities that are hardly accessible in
analytic calculations~-- one of the most prominent and, as it will turn out in the
following, most relevant system-specific quantities is the density (or number) of states
with energy $E$, $g(E)$~-- opens new perspectives for the physical interpretation and classification
of cooperative processes such as phase transitions. This is particularly interesting for small systems,
where conventional statistical analyses are often little systematic and a general
concept seems to be missing. This is apparently reflected in conformational studies in 
the biosciences, where often novel terminologies are invented for basically the same
classes of transitions. The introduction of a unifying scheme appears insofar difficult as
finite-size effects in transitions of small systems influence the thermal fluctuations of
transition indicators like order parameters. Maximum energetic fluctuations, represented
by peaks in the specific heat, do not necessarily coincide with peak temperatures of
fluctuations of structural quantities such as the radius of gyration~\cite{bj0,bj2}. For extremely large
systems, which are well described by the theoretical concept of the ``thermodynamic limit'',
this \emph{canonical} approach is appealing as the fluctuation peaks scale with system size and
finally collapse at the same temperature,
allowing for the definition of a unique transition point. However, for many systems, in particular
biomolecules, the assumption of the thermodynamic limit is nonsensical and the explicit 
consideration and understanding of finite-size effects is relevant. 

\section{Microcanonical vs.\ canonical temperature}
\label{temp}

The \emph{microcanonical} analysis~\cite{gross1} allows for such an in-depth analysis of smallness effects.
It is completely based on the entropy as a function of the system energy, which is related
to the density of states via $S(E)=k_{\rm B}\ln\, g(E)$, where $k_{\rm B}$ is the Boltzmann
constant. A major advantage of this quantity is the possibility to introduce the 
temperature as a \emph{derived} quantity via
\begin{equation}
\label{eq:TE}
T(E)=\left[\frac{\partial S(E)}{\partial E}\right]^{-1}
\end{equation}
which is commonly refered to as the ``microcanonical temperature''. This
terminology is misleading 
since $S(E)$ and thus $T(E)$ do not depend on the choice of any statistical
ensemble associated to a certain thermal 
environment of the system. Thus, the physical meaning of $S(E)$ and $T(E)$ 
is not restricted to systems
well-described by the microcanonical ensemble only (i.e., for systems with
constant energy). The introduction of the temperature via 
Eq.~(\ref{eq:TE})  
is also useful for another reason. It applies independently of the system size
and it is not coupled 
to any equilibrium condition.
To make use of theoretical concepts like the thermodynamic limit or 
quasiadiabatic process flow is not necessary.

The typically used heatbath concept for introducing the temperature in the context of the
thermodynamic equilibrium of heatbath and system is useful for large systems when
thermal fluctuations become less relevant and finite-size effects disappear. However, as it has already
been known from the early days of statistical mechanics, the statistical ensembles turn over
to the microcanonical ensemble in the thermodynamic limit. This is easily seen for the example
of the fluctuations about the mean energy in the canonical ensemble. In this case, the canonical
statistical partition 
function, linked to the free energy $F(T^{\rm can}_{\rm system})$, can be written as an integral
over the energy space,
\begin{equation}
\label{eq:Z}
Z(T^{\rm can}_{\rm system})=\int_{E_{\rm min}}^{E_{\rm max}} dE g(E) e^{-E/k_{\rm B}T^{\rm can}_{\rm system}}
=e^{F/k_{\rm B}T^{\rm can}_{\rm system}},
\end{equation}
where the canonical system temperature $T^{\rm can}_{\rm system}$ 
corresponds in equilibrium to the heatbath temperature 
$T_{\rm heatbath}$ (which is an adjustable external thermal control parameter in experiments): 
\begin{equation}
\label{eq:equ}
T^{\rm can}_{\rm system}\equiv T_{\rm heatbath}.
\end{equation}
Therefore, in equilibrium, the mean energy at a given heatbath temperature can simply be calculated as:
\begin{equation}
\label{eq:meanE}
\langle E\rangle(T_{\rm heatbath})=\frac{1}{Z(T_{\rm heatbath})}\int dE\, E\,g(E)e^{-E/k_{\rm B}T_{\rm heatbath}}.
\end{equation}
The heat capacity 
$C_V(T_{\rm heatbath})=d\langle E\rangle/dT_{\rm heatbath}=(\langle E^2\rangle-\langle E\rangle^2)/k_{\rm B}T_{\rm heatbath}^2$ 
must always be 
nonnegative because of the thermodynamic stability of matter, i.e., $\langle E\rangle$ increases
monotonously with $T_{\rm heatbath}$. For this reason, the dependence of $\langle E\rangle$ 
on the temperature can trivially be inverted,  
$T^{\rm can}_{\rm system}(\langle E\rangle)$,
where we have made use of the equilibrium condition~(\ref{eq:equ}). In complete analogy to the microcanonical
definition of the temperature in Eq.~(\ref{eq:TE}), we introduce the canonical entropy via the relation
\begin{equation}
\label{eq:Tcan}
T^{\rm can}_{\rm system}(\langle E\rangle)= 
\left[\frac{\partial S^{\rm can}(\langle E\rangle)}{\partial \langle E\rangle}\right]_{N,V}^{-1},
\end{equation}
where particle number $N$ and volume $V$ are kept constant. The canonical entropy can explicitly
be expressed by the celebrated equation that links thermodynamics and statistical mechanics:
\begin{equation}
\label{eq:S}
S^{\rm can}(\langle E\rangle)=\frac{1}{T^{\rm can}_{\rm system}(\langle E\rangle)}
\left[\langle E\rangle - F(T^{\rm can}_{\rm system}(\langle E\rangle))\right].
\end{equation}
Instead of considering the canonical ensemble by fixing $T_{\rm heatbath}$, $V$, and $N$ as external 
parameters, we have turned to the caloric representation, where $\langle E\rangle$, $V$, and $N$ are
treated as independent variables. If the fluctuations of energy about the mean
value $\langle E\rangle$ vanish,
the canonical ensemble thus corresponds to the microcanonical ensemble. This is obvious in the 
thermodynamic limit, where the relative width of the canonical energy distribution,
$\Delta E/\langle E\rangle=\sqrt{\langle E^2\rangle-\langle E\rangle^2}/\langle
E\rangle=\sqrt{k_{\rm B}T_{\rm heatbath}^2C_V(T_{\rm heatbath})}/\langle
E\rangle\propto 1/\sqrt{N}$ vanishes,
$\lim_{N\to\infty}(\Delta E/\langle E\rangle) = 0$, since $C_V$ and $\langle E\rangle$ are extensive 
variables as they scale with $N$. Thus, $\langle E\rangle = E$ 
and $T(E)=T^{\rm can}_{\rm system}(\langle E\rangle)$ in the thermodynamic limit.

\section{Small systems}
\label{small}

Microcanonical and canonical temperatures do typically not coincide, if finite-size effects
matter. This is particularly apparent under conditions where cooperative changes of macrostates, 
such as conformational transitions of finite molecular systems, occur. In transitions with
structure formation, the conformational entropies associated to volume and
surface effects in the formation of compact states of the system
compete with the energetic differences of particles located at the surface or in the interior
of the structure.
Examples for such morphologies of finite size 
are atomic clusters, spin clusters, the interface of demixed fluids, 
globular polymers or proteins, crystals, etc. Since many interesting systems such as heterogeneous
biomolecules like proteins are ``small'' in a sense that a thermodynamic limit does not exist
at all, it is useful to build up the analysis of transitions of small systems on the most general
grounds. These are, as we have argued above, best prepared by the microcanonical or caloric
approach. 

The folding of a protein is an example for a subtle structure formation process of a finite system,
where effects on nanoscopic scales (e.g., hydrogen bonds being responsible for local secondary structures
such as helices and sheets) and cooperative behavior on mesoscopic scales (such as the less well-defined
hydrophobic effect which primarily drives the formation of the global, tertiary structures) contribute
to the stable assembly of the native fold which is connected to the biological function of the protein.
Proteins are linear polypeptides, i.e., they are composed of amino acids linked by peptide bonds.
There are twenty types of amino acids that have been identified in bioproteins, all of them 
differ in their chemical composition and thus possess substantial differences in their
physical properties and chemical reactivity. The mechanism of folding depends on the sequence
of amino acids, not only the content, i.e., a protein is a disordered system. It is one of the central
questions, which mutations of a given amino-acid sequence can lead to a relevant change of morphology
and thus the loss of functionality. The atomistic interaction types, scales, and ranges are 
different as well. For this reason, in compact folds, frustration effects may occur. Disorder and
frustration cause glassy behavior, but how glassy is a single protein and can this be generalized?

Since details seem to be relevant for folding, it has commonly been believed
that the folding of a certain protein is a highly individual process. Thus, it was a rather surprising
discovery that the search for a stable fold can be a cooperative one-step process which can qualitatively
and quantitatively be understood by means of the statistical analysis of a single effective, mesoscopic 
order parameter. The free-energy landscape turned out to be very simple (exhibiting only a single barrier
between folded and unfolded conformations) for this class of 
``two-state folders''~\cite{fersht1,fersht2}. Of course, folding pathways for other proteins can be more complex as also intermediate
states can occur~\cite{sbj1,sbj2}. However, this raises the question about cooperativity and
the generalization of folding behavior in terms of conformational transitions similar to
phase transitions in other fields of statistical mechanics. In the following, we are going to
discuss a structure formation process, the aggregation of a finite system of
heteropolymers, by a general
microcanonical approach in order to show how the conformational transition behavior of a small system is indeed
related to thermodynamic phase transitions.

\section{Exemplified nucleation process: Aggregation of proteins}
\label{first}

As an example for the occurrence of hierarchies of subphase transitions accompanying 
a nucleation process, we are going to discuss molecular aggregation~\cite{jbj1,jbj2,jbj3} 
by means of a simple coarse-grained hydrophobic-polar heteropolymer model~\cite{still1,baj1}. 
In this so-called AB model, only hydrophobic (A) and hydrophilic (B) monomers line up in linear 
heteropolymer sequences. In the following, we consider the aggregation of four chains with
13 monomers~\cite{jbj2}. All chains have the same Fibonacci sequence $AB_2\-AB_2\-ABAB_2\-AB$~\cite{still1}. Folding and aggregation
of this heteropolymer have already been subject
of former studies~\cite{jbj1,jbj2}. 

In the model used here, bonds between adjacent monomers have fixed length unity. 
Nonbonded monomers of individual chains but likewise mono\-mers of different chains
interact pairwisely via Lennard-Jones-like potentials. The explicit form
of the potentials depends on the types of the
interacting mono\-mers. Pairs of hydrophobic and unlike mono\-mers attract, pairs of 
polar mono\-mers repulse
each other. This effectively accounts for the hydrophobic-core formation of proteins in 
polar solvent. Details of our aggregation model and of the implementation of
the multicanonical 
Monte Carlo simulation method are described in Ref.~\cite{jbj2}.

\begin{figure}[t!]
\centerline{\includegraphics[width=90mm]{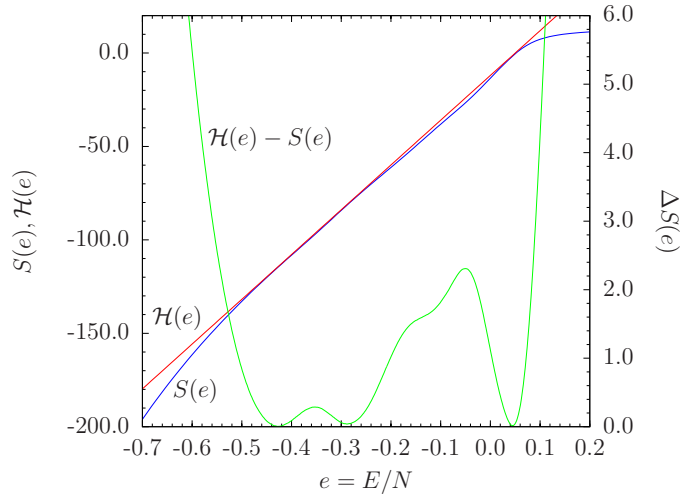}}
\caption{\label{fig:mic} 
Microcanonical entropy $S(e)$, the Gibbs hull ${\cal H}(e)$, and the deviation
$\Delta S(e)={\cal H}(e)-S(e)$ as functions of the energy per monomer.}
\end{figure}

The multicanonical computer simulations enabled us to obtain a precise estimate for the 
microcanonical entropy $S(E)$ of the multiple-chain system~\cite{jbj2}, as shown
in Fig.~\ref{fig:mic} as a function of the energy per monomer, $e=E/N$.
The entropy curve
is convex in the energetic aggregation transition region as expected for a 
first-order-like nucleation
transition of a finite system~\cite{gross1}. 
The Gibbs tangent ${\cal H}(e)$, connecting the two coexistence 
points where concave and
convex behavior change, provides the least possible 
overall concave shape of $S(e)$ in this region.
The difference between the Gibbs hull and the entropy curve,
$\Delta S(e)={\cal H}(e)-S(e)$, is also shown in Fig.~\ref{fig:mic}. Not only the 
entropic suppression in the transition region is clearly visible, it is also
apparent that the transition possesses an internal structure. 

\begin{figure}[t!]
\centerline{\includegraphics[width=90mm]{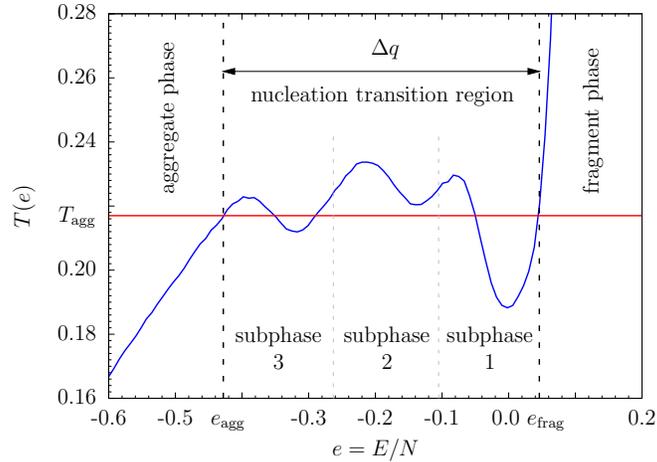}}
\caption{\label{fig:tcal} 
Microcanonical temperature $T(e)$ as a function of energy per monomer, $e=E/N$, where $N$ is the
total number of all monomers in the system. The horizontal Maxwell line marks the aggregation temperature
$T_{\rm agg}$, obtained by the Gibbs construction. Vertical dashed lines separate the different
phases and subphases, respectively.}
\end{figure}

In order to
better understand the subphases, we discuss in the following
the inverse derivative of the entropy, the microcanonical 
temperature~(\ref{eq:TE}), $T(e)$, which is plotted in Fig.~\ref{fig:tcal}. 
The slope of
the Gibbs tangent corresponds to the Maxwell line in Fig.~\ref{fig:tcal} at the
aggregation temperature $T_{\rm agg}\approx 0.217$. Therefore, the intersection points
of the Maxwell line and the 
temperature curve define the energetic phase boundaries $e_{\rm agg}$ and $e_{\rm frag}$,
respectively, as both phases coexist at $T_{\rm agg}$. 

For energies $e<e_{\rm agg}\approx -0.43$, 
conformations of a single aggregate, composed of all four chains, dominate. On the
other hand, conformations with $e>e_{\rm frag}\approx 0.05$ are mainly entirely fragmented, i.e.,
all chains can form individual conformations, almost independently of each other.
The entropy is governed by the contributions of the individual translational entropies
of the chains, outperforming the conformational entropies. The translational entropies
are only limited by the volume which corresponds to the simulation box size.
The energetic difference $\Delta q=e_{\rm frag}-e_{\rm agg}$, serving as an estimator
for the latent heat, is obviously larger
than zero, $\Delta q \approx 0.48$. It thus corresponds to the total energy necessary
to entirely melt
the aggregate into fragments at the aggregation (or melting) temperature.

\begin{figure}[t!]
\centerline{\includegraphics[width=90mm]{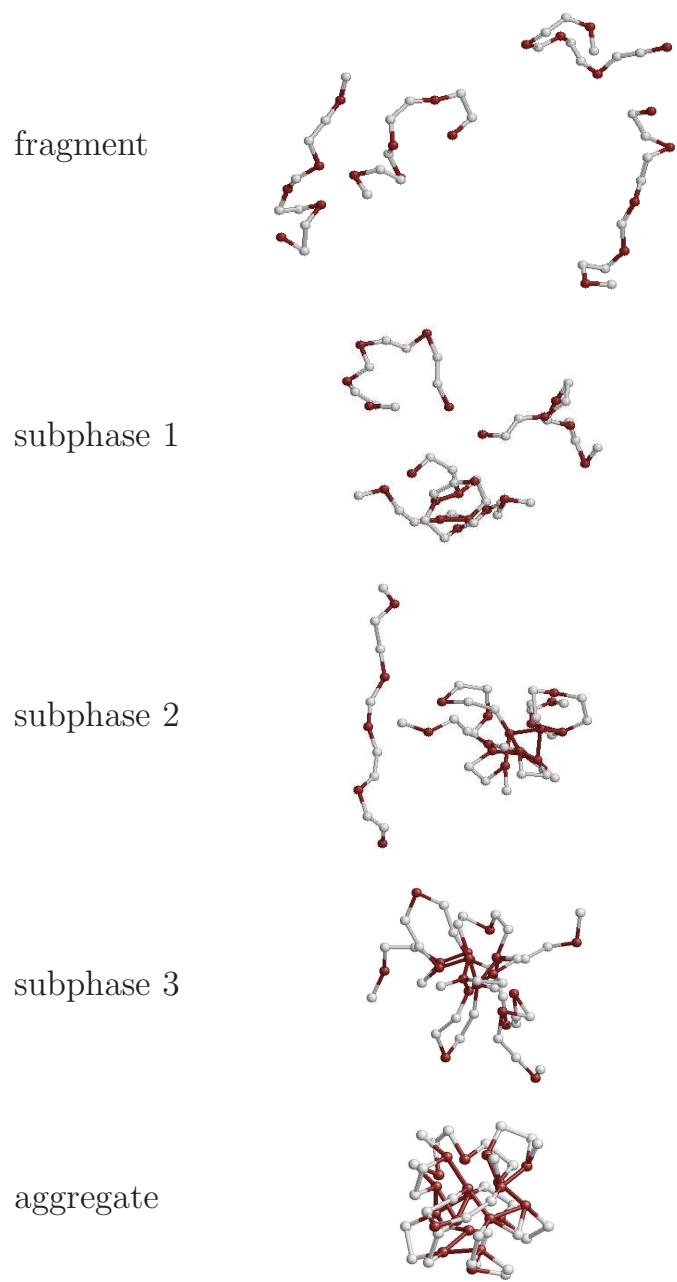}}
\caption{\label{fig:pics} 
Representative conformations in the different structural phases.
}
\end{figure}

The relation between energy and microcanonical temperature in the aggregate phase
and in the fragment 
phase is as intuitively expected: 
With increasing energy, also $T(e)$ increases. However, much more interesting is the behavior
of the system in the energy interval $e_{\rm agg}<e<e_{\rm frag}$, which represents the
energetic nucleation transition region. Figure~\ref{fig:tcal}
clearly shows that in our example 
$T(e)$ changes the monotonic behavior three times in this
regime. Representative conformations in the diffferent structural phases
  are shown in Fig.~\ref{fig:pics}.
The change of monotony of the microcanonical temperature curve is called backbending effect,
because the temperature decreases while energy is increased. This rather little intuitive
behavior is a consequence of the suppression of entropy in this regime 
($S(E)$ is convex in the backbending region). The surface entropy per monomer
vanishes in the thermodynamic limit~\cite{jbj2}.

If two chains aggregate (subphase 1 in Fig.~\ref{fig:tcal}), the total translational entropy
of the individual chains is reduced by $k_B\ln\, V$, where $V$ is the volume (corresponding
to the simulation box size), whereas the energy of the aggregate is much smaller than the 
total energy of the system with the individual chains separated. Thus, the energy 
associated to the interaction between different chains, i.e., the cooperative
formation of inter-chain contacts between monomers of different chains, is highly
relevant here. This causes the latent heat between the completely 
fragmented and the two-chain aggregate phase to be nonzero which signals a first-order-like 
transition. This procedure continues when an additional chain joins a two-chain cluster.
Energetically, the system enters subphase 2. Qualitatively, the energetic and entropic reasons for
the transition into this subphase are the same as explained before, since it is the same
kind of nucleation process. In our example of four chains interacting with each other,
there is another subphase 3 which also shows the described behavior. The energetic width
of each of the subphase transitions corresponds to the respective latent heat gained by
adding another chain to the already formed cluster. The subphase boundaries
(vertical dashed, gray lines 
in Fig.~\ref{fig:tcal}) have been defined to be the inflection points in the raising 
temperature branches, thus enclosing a complete ``oscillation'' of the temperature as a 
function of energy. The energetic subphase transition points are located at
$e_{\rm 12}\approx -0.11$ and $e_{\rm 23}\approx -0.26$, respectively. Therefore, the latent
heat associated to these subphase transitions is in all three cases about 
$\Delta q_{ij}\approx 0.16$ ($i,j=1,2,3$, $i\neq j$), thus being one third of the total 
latent heat of the complete nucleation process. This reflects the high systematics of
subphase transitions in first-order nucleation processes. 

\section{Summary}

The most interesting result from this heteropolymer aggregation study is that first-order phase
transitions such as nucleation processes can be understood as a composite of hierarchical
subphase transitions, each of which exhibits features of first-order-like transitions.
Since with increasing number of chains the microcanonical entropy per chain 
converges to the Gibbs hull in the transition region, the ``amplitudes'' of the backbending
oscillations and the individual latent heats of the subphases become smaller and
smaller~\cite{jbj2,jbj3}. 
Thus, in the thermodynamic limit, the heteropolymer aggregation transition is
a first-order nucleation
process
composed of an infinite number of infinitesimally ``weak'' first-order-like
subphase transitions.

\section*{Acknowledgments}

Supercomputer time has been provided by
the Forschungszentrum J\"ulich under Project Nos.\ hlz11, jiff39, and jiff43.




\end{document}